# Did We Miss Something Important? Studying and Exploring Variable-Aware Log Abstraction

Zhenhao Li*, Chuan Luo†, Tse-Hsun (Peter) Chen*, Weiyi Shang*, Shilin He†, Qingwei Lin†, and Dongmei Zhang†
*Department of Computer Science and Software Engineering, Concordia University, Montreal, Canada
{l_zhenha, peterc, shang}@encs.concordia.ca
†Microsoft Research, China
{chuan.luo, shilin.he, qlin, dongmeiz}@microsoft.com

*Abstract*—Due to the sheer size of software logs, developers rely on automated techniques for log analysis. One of the first and most important steps of automated log analysis is log abstraction, which parses the raw logs into a structured format. Prior log abstraction techniques aim to identify and abstract all the dynamic variables in logs and output a static log template for automated log analysis. However, these abstracted dynamic variables may also contain important information that is useful to different tasks in log analysis. In this paper, we investigate the characteristics of dynamic variables and their importance in practice, and explore the potential of a variable-aware log abstraction technique. Through manual investigations and surveys with practitioners, we find that different categories of dynamic variables record various information that can be important depending on the given tasks, the distinction of dynamic variables in log abstraction can further assist in log analysis. We then propose a deep learning based log abstraction approach, named VALB, which can identify different categories of dynamic variables and preserve the value of specified categories of dynamic variables along with the log templates (i.e., variable-aware log abstraction). Through the evaluation on a widely used log abstraction benchmark, we find that VALB outperforms other state-of-the-art log abstraction techniques on general log abstraction (i.e., when abstracting all the dynamic variables) and also achieves a high variable-aware log abstraction accuracy that further identifies the category of the dynamic variables. Our study highlights the potential of leveraging the important information recorded in the dynamic variables to further improve the process of log analysis.

## I. INTRODUCTION

Logs play an important role in software systems to record system execution behaviors. Practitioners leverage logs to assist in various tasks in the process of software development and maintenance, such as failure diagnosis [1], [2], [3], [4], [5], [6], program comprehension [7], [8], [9], [10], [11], [12], [13], and anomaly detection [14], [15], [16], [17]. Although logs contain rich system run-time information, there are challenges associated with log analysis [18]. For example, modern software systems generate a large number of logs on a daily basis, resulting in tens of gigabytes or even terabytes of data [19], [20], [21]. Therefore, developers usually rely on automated techniques for log analysis.

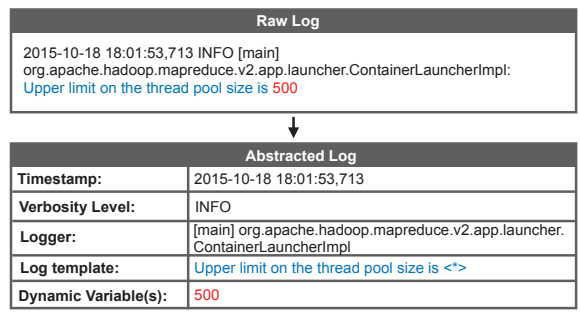

Fig. 1. An example of the log abstraction process.

To effectively facilitate automated log analysis, log abstraction (also called log parsing) techniques [14], [19], [20], [22], [23] are used to process the raw logs into a more structured format. Figure 1 shows an example of the log abstraction process. The raw log in the example is composed of a message header (e.g., timestamp) that can be configured via the logging library, the static words that remain constant, and the dynamic variables that may vary depending on the run-time behaviors. The goal of log abstraction techniques is to identify the static log templates and abstract the dynamic variables from the raw logs and output the log templates. The sequence of log templates can then be leveraged for further log analysis (e.g., anomaly detection [24], [17]).

Prior studies propose various log abstraction techniques using different algorithms [19], such as frequent pattern mining (e.g., Logram [20] and LFA [25]), clustering algorithms (e.g., SHISO [26] and Lenma [27]), heuristics (e.g., Drain [22], AEL [23], and IPLoM [28]), and combined approaches (e.g., ULP [29]). Their common goal is to abstract all dynamic parts of logs and output the remaining static words. However, the recorded dynamic values can provide valuable information to assist log understanding and analysis. In this paper, we aim to understand the characteristics of dynamic variables and their importance in log analysis. We then seek to propose a log abstraction approach that can selectively abstract dynamic variables that belong to specific categories based on the needs.

We first empirically study the dynamic variables in logs. We manually study a widely used log abstraction benchmark data set [19] to uncover the characteristics of dynamic variables in logs and uncover 10 categories of dynamic variables. We then have a questionnaire survey with practitioners at Microsoft to investigate how do practitioners in the industry consider the usage and importance of dynamic variables in logs. Through our empirical study and survey, we find that different categories of dynamic variables record valuable information that can be important for different tasks. In our survey with industry practitioners, we also find that the practitioners acknowledge the importance of dynamic variables in logs, and a log abstraction technique that can preserve the categories of dynamic variables as specified may further help log analysis.

Motivated by our findings and practitioners' feedback, we then propose VALB, which is a **V**ariable-**A**ware **L**og a**B**straction approach that can identify the static and dynamic parts in logs (as the prior log abstraction techniques do), and also further identify the categories of dynamic variables. **Practitioners can specify the categories of dynamic variables based on their needs, and the values of such dynamic variables will be preserved along with the log templates**. Overall, VALB achieves an average accuracy of 96.1% for general log abstraction, which is better than other state-of-the-art log abstraction techniques (average accuracy ranges from 74.5% to 82.5%). VALB also achieves an average accuracy of 95.5% for variable-aware log abstraction that further considers the correctness of identifying different categories of dynamic variables. Moreover, we find that the models of VALB are still efficient on a new system by re-using the models trained from other systems and fine-tune the models with a small amount of logs in the new system.

The contributions of this paper are as follows[1]

- We investigate the characteristics and importance of dynamic variables in logs, which are omitted by prior log abstraction techniques. We find that different categories of dynamic variables record valuable information that can be important for different tasks, point out the need of a variable-aware log abstraction technique.
- We propose a deep learning approach, VALB, which is the first log abstraction approach that can further identify the categories of dynamic variables in the process of log abstraction. VALB achieves promising results in variable-aware log abstraction and also outperforms prior state-of-the-art techniques in general log abstraction.
- We explore the potential of variable-aware log abstraction on assisting in log-based downstream tasks and find that variable-aware log abstraction can be leveraged to improve the performance of anomaly detection.

In short, our study uncovers the importance of dynamic variables and highlights future research opportunities on studying the potential of leveraging dynamic variables in logs to further assist in log analysis.

**Paper Organization.** Section II discusses the motivating examples of our study. Section III presents our empirical study on dynamic variables in logs. Section IV describes the data preparation and the deep learning framework of VALB. Section V evaluates VALB by proposing and answering three research questions. Section VI discusses the potential of variable-aware log abstractions on assisting in log-based downstream tasks. Section VII discusses the threats to the validity of our study. Section VIII summarizes the related works. Section IX concludes the paper.

[1] We share the data of this paper in the repository: https://github.com/ginolzh/Variable_Aware_Log_Abstraction.

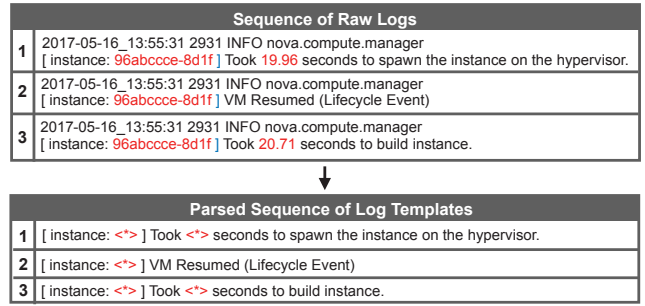

Fig. 2. An example of log parsing before analyzing the sequences of logs. The dynamic variables in the raw logs and abstracted variables in the parsed log templates are marked in red.

## II. MOTIVATING EXAMPLES

Log abstraction has shown to be a crucial first step towards further log analysis [19], [14]. Prior log abstraction techniques aim at abstracting all the dynamic variables and output a static log template [19]. However, during our collaboration with industry, practitioners mention that such log abstraction process will result in the loss of important information recorded by dynamic variables. They consider that different categories of dynamic variables record different information, which can be important to specific tasks.

We take anomaly detection as an example. Anomaly detection tools [15], [16] analyze the sequences of log templates generated from log abstraction techniques to detect system anomalies. Since the dynamic variables have already been abstracted, the input log templates only contain the information of static words in logs. Figure 2 shows an example of the process of log abstraction on a sequence of logs. We take the example logs from OpenStack, an open-source anomaly detection data set provided by the *LogPAI* project [30]. The data sets provide sequences of logs that are generated from normal and abnormal (e.g., system crash) system execution behaviors. In the example shown in Figure 2, the raw logs record a series of run-time information of an instance (i.e., a node) "96abcccce-8d1f". The dynamic variables (i.e., instance ID and seconds taken by an action) are then abstracted as wildcards in the output sequences of log templates.

The anomaly detection approach then learns from the sequences of log templates in the training data, and predicts whether a given log sequence indicates a normal or abnormal run-time behavior. However, the dynamic variables may also play an important role in distinguishing normal and abnormal system behaviors. For example, in the log template *"[instance: *] Took * seconds to spawn the instance on the hypervisor"*, the average value of *"seconds"* in normal log sequences is

19.78, while the average value in abnormal sequences is 53.87. For the other log template *"[instance: *] Took * seconds to build instance."*, the average values of *"seconds"* are 20.63 and 39.21 in normal and abnormal log sequences, respectively. This indicates that this anomaly might be a performance issue due to network latency.

Therefore, apart from the static parts in the logs sequences, dynamic variables may also contain important information for log analysis, yet such information is usually abstracted by prior log abstraction techniques. An investigation of the characteristics of dynamic variables and their importance in practice may further help improve log abstraction techniques and to better assist in log analysis.

## III. STUDYING THE DYNAMIC VARIABLES IN LOGS

Motivated by practitioners' awareness of the importance of dynamic variables, in this section, we study the dynamic variables in logs. We first conduct a manual study on a widely used log data set [30] to study the characteristics of the dynamic variables and categorize the information they record. We then conduct a questionnaire survey [31], [32] at Microsoft to investigate how do practitioners consider the role that dynamic variables play in log analysis.

### A. Manually Studying and Characterizing the Dynamic Variables in Logs

**Studied Data Sets.** We use the log data sets and benchmarks provided by *LogPAI* [30] to study the characteristics of dynamic variables in logs. LogPAI includes 16 data sets of logs generated from both open-source and commercial systems across various domains, and is widely used in prior log analysis studies [16], [15]. Each data set also provides a subset of 2,000 logs with manually derived log templates as the ground truth for evaluating the accuracy of log abstraction techniques. The manually derived log templates are commonly used in prior log abstraction studies for evaluating log abstraction accuracy [19], [20], [22]. We call this subset of data sets *log abstraction benchmark data sets* in the rest of our paper.

In our manual analysis, we study the above mentioned log abstraction benchmark data sets (i.e., 16 data sets in total, 2000 logs per data set, each log is manually labeled with its corresponding log template for the evaluation of log abstraction). Table I presents the details of the log abstraction benchmark data sets, including the number of log templates (NOL), number of log templates that have abstracted dynamic variables with the percentage among the total number of log templates (TWV (%)), and the number of abstracted dynamic variables (NOV). The number of log templates ranges from 6 (Apache) to 341 (Mac), and the number of abstracted dynamic variables ranges from 8 (Apache) to 595 (Mac). We also find that, in each data set, more than half of the log templates have abstracted dynamic variables (i.e., not pure static messages).

**Manual Investigation on Dynamic Variables.** In order to investigate what kind of information do those dynamic variables record and their potential importance, we manually study the logs in the log abstraction benchmark data sets to uncover the characteristics of dynamic variables in logs. Similar to the process of manual study in prior studies [33], [34], [35], [36], our study involves the following steps:

TABLE I
AN OVERVIEW OF THE LOG ABSTRACTION BENCHMARK DATA SETS. NOL: NUMBER OF LOG TEMPLATES, TWV (%): PERCENTAGE OF LOG TEMPLATES WITH VARIABLES, NOV: NUMBER OF VARIABLES

| Name | NOL | TWV (%) | NOV |
|---|---|---|---|
| Android | 166 | 107 (64.5%) | 320 |
| Apache | 6 | 6 (100.0%) | 8 |
| BGL | 120 | 106 (88.3%) | 269 |
| Hadoop | 114 | 90 (78.9%) | 160 |
| HDFS | 14 | 14 (100.0%) | 36 |
| HealthAPP | 75 | 44 (58.7%) | 74 |
| HPC | 46 | 27 (58.7%) | 52 |
| Linux | 118 | 67 (56.8%) | 117 |
| Mac | 341 | 268 (78.6%) | 595 |
| OpenSSH | 27 | 23 (85.2%) | 56 |
| OpenStack | 43 | 42 (97.7%) | 91 |
| Proxifier | 8 | 8 (100.0%) | 23 |
| Spark | 36 | 24 (66.7%) | 39 |
| Thunderbird | 149 | 97 (65.1%) | 202 |
| Windows | 50 | 31 (62.0%) | 66 |
| Zookeeper | 50 | 38 (76.0%) | 80 |
| *Total* | 1363 | 992 (72.8%) | 2188 |

*Step 1*: We go through all the log templates that have abstracted dynamic variables in the data sets (992 log templates in total). For the abstracted dynamic variables in each log template, the first author of this paper (i.e., A1) checks its corresponding original logs and note down what kind of information is recorded by each dynamic variable.

*Step 2*: Two authors of this paper (i.e., A1 and A2) independently re-visit the notes in Step 1 and derive a list of categories that can describe the characteristics of the dynamic variables [37]. A1 and A2 then discuss the derived results. The categories are refined and revised during the discussion in this step.

*Step 3*: A1 and A2 use the categories derived in Step 2 to label every dynamic variable in the data sets (2,188 dynamic variables in total) independently. The results between the two authors have a Cohen's Kappa of 0.79, which is a substantial level of agreement [38]. We discuss the disagreements until a consensus is reached. We then compute the number of dynamic variables that belongs to each category.

**Categories of Dynamic Variables.** In our manual investigation, we uncover 10 categories of dynamic variables. Table II presents each category with its abbreviation (**Abbrev.**), description, example, and the number of variables that belong to this category. The dynamic variables in the example are marked in bold and underline.

Overall, we find that Object ID (OID), Location Indicator (LOI), Object Amount (OBA), and Object Name (OBN) have a relatively higher presence in our studied dynamic variables (255 to 739, out of 2,188). Among them, Object ID (OID) records the identification information of an object, such as a session ID or user ID. Location Indicator (LOI) shows the location information of an object. It can be path information, a URI, or an IP address. Object Name (OBN) shows the name of an object (e.g., domain name, task name, job name), and Object Amount (OBA) records the amount of an object (e.g.,

TABLE II
THE MANUALLY-DERIVED CATEGORIES OF DYNAMIC VARIABLES WITH THEIR CORRESPONDING ABBREVIATIONS (**Abbrev.**). DYNAMIC VARIABLE IN THE EXAMPLE IS MARKED IN **BOLD AND UNDERLINE**

| Name | Abbrev. | Description | Example | Total Number |
|---|---|---|---|---|
| Object ID | OID | Identification information of an object. | Added attempt **1445144423722** to list of failed maps. | 739/2,188 |
| Location Indicator | LOI | Location information of an object. | Adding path spec: **/mapreduce**. | 419/2,188 |
| Object Name | OBN | Name of an object. | ServerFileSystem domain **root10-local** is full. | 255/2,188 |
| Type Indicator | TID | Type information of an object or an action. | Using configuration type **1**. | 27/2,188 |
| Switch Indicator | SID | Status of a switch variable. | Saw change in network reachability (isReachable= **2**). | 7/2,188 |
| Time/Duration of an Action | TDA | Time or duration of an action. | Scheduled snapshot period at **10** second(s). | 137/2,188 |
| Computing Resources | CRS | Information of computing resource. | Combo kernel: **126MB** LOWMEM available. | 143/2,188 |
| Object Amount | OBA | Amount of an object. | Total of **23** ddr error(s) detected and corrected. | 340/2,188 |
| Status Code | STC | Status code of an object or an action. | mod-jk child workerEnv in error state **7**. | 93/2,188 |
| Other Parameters | OTP | Other information does not belong to the above categories. | calvisitor kernel: payload Data **0700** to list of failed maps. | 28/2,188 |

TABLE III
THE SURVEY RESULTS OF "OPINIONS ON THE CATEGORIES OF DYNAMIC VARIABLES IN LOGS". UI: USUALLY IMPORTANT, CBI: CAN BE IMPORTANT IN SOME SITUATIONS, UNI: USUALLY NOT IMPORTANT.

| Name | UI | CBI | UNI |
|---|---|---|---|
| Object ID (OID) | **45.5%** | 40.9% | 13.6% |
| Location Indicator (LOI) | **50.0%** | 40.9% | 9.1% |
| Object Name (OBN) | **59.1%** | 31.8% | 9.1% |
| Type Indicator (TID) | **63.6%** | 27.3% | 9.1% |
| Switch Indicator (SID) | **54.5%** | 27.3% | 18.2% |
| Time/Duration of an Action (TDA) | 31.8% | **63.6%** | 4.6% |
| Computing Resources (CRS) | **45.5%** | 36.4% | 18.1% |
| Object Amount (OBA) | **40.9%** | 40.9% | 18.2% |
| Status Code (STC) | **68.2%** | 31.8% | 0.0% |
| Other Parameters (OTP) | 9.1% | **63.6%** | 27.3% |
| *Average* | 46.8% | 40.5% | 12.7% |

number of errors, number of nodes).

Time or Duration of an Action (TDA) and Computing Resources (CRS) also have over 100 dynamic variables in our studied data sets. TDA shows the time that an action happens or the duration of an action, and CRS shows how many computing resources are in use or left (e.g., memory or disk space). For the remaining four categories (i.e., Type Indicator (TID), Switch Indicator (SID), Status Code (STC), and Other Parameters (OTP)), they have relatively fewer numbers among the studied dynamic variables. However, they can also be important in some situations. For example, Status Code (STC) can show the code for some crucial events (e.g., an error code), which is usually important for error diagnosis.

*B. A Survey on Log Analysis and Dynamic Variables*

In our manual study, we find that different categories of dynamic variables in logs record various system run-time behaviors and uncover 10 categories of dynamic variables. In order to investigate how do developers in the industry consider the usage of dynamic variables in logs when doing log analysis, we conduct a questionnaire survey in Microsoft. Specifically, we conduct the survey in several production teams in Microsoft with more than one hundred full-time engineers in total. We first have a pilot survey with three engineers to collect their feedback. We make minor modifications to revise the overall presentation of the survey based on the received feedback. The final version of the survey has 17 questions, divided into four parts. We then distribute the survey link to the group chat of those teams. We receive 22 responses in total, the participants are engineers from various production teams such as cloud computing and social network services. The roles of our survey participants include development, quality assurance, and product management. Based on their survey responses, we find that the participants are involved in various tasks related to log analysis such as root cause analysis and incident monitoring. Below, we discuss each part of survey questions in detail.

**Experience of the Participants (Q1).** We ask the participants how many years of experience do they have in software development and maintenance. On average, the participants have 6 years of experience.

**Opinions on the Categories of Dynamic Variables in Logs (Q2–Q11).** We provide the participants all the 10 categories of dynamic variables that we uncovered with a corresponding example and ask the participants to consider the importance of each category in practice. Table III present the results of the participants' opinions on the categories of dynamic variables. For each category, participants can select if the category is "Usually important (UI)", "Can be important in some situations (CBI)", or "Usually not important (UNI)". The highest number for each category is marked in **bold**.

Overall, most of the participants consider the dynamic variables are usually important, or can be important in some situations. For all the 10 categories, from 27.3% to 63.6% of the participants consider that they can be important in some situations. For 5 out of the 10 categories (LOI, OBN, TID, SID, and STC), more than half of the participants consider that they are usually important. The results show that developers acknowledge the importance of dynamic variables in logs, while some variables are usually important and some are important depending on the situations.

**Follow-up Questions on Dynamic Variables and Log Analysis (Q12-Q16).** In this part, we ask the participants five multiple-choice questions related to the dynamic variables in logs and log analysis. For each question, participants can choose one score from 1 to 5, where 1 represents for "Very low extent", and 5 represents for "Very high extent". Table IV presents the survey questions and results. The column of *Avg.* and *Med.* shows the average and median score, respectively.

Overall, the average and median scores for Q12 - Q16 are all above 4.0. The results of Q12 (an average of 4.5 and a median of 5.0) show that the participants acknowledge the importance of dynamic variables in log analysis. The results of Q14 (an average of 4.2 and a median of 4.0) then show that developers believe further distinguishing the categories of dynamic variables may further help log analysis. Based on the results of Q13 (an average and a median of 4.0), the participants consider that our derived categories can represent the dynamic variables to a high extent. Combining the results of Q15 and the percentage of CBI in Table III, participants consider that different categories of dynamic variables may play different roles in log analysis, depending on the specific tasks or requirements. In Q16, most of the participants consider that it will be more helpful if the log abstraction technique can further identify and keep some certain categories of dynamic variables during the log abstraction process.

TABLE IV
LIST OF QUESTIONS AND RESULTS FOR "FOLLOW-UP QUESTIONS ON DYNAMIC VARIABLES AND LOG ANALYSIS", THE ANSWERS ARE IN A SCALE FROM 1 (VERY LOW EXTENT) TO 5 (VERY HIGH EXTENT).

| | Question | Avg. | Med. |
|---|---|---|---|
| Q12 | To what extent do you think the dynamic variables are important for log analysis? | 4.5 | 5.0 |
| Q13 | For the 10 categories of dynamic variables, to what extent do you think they can represent the dynamic variables in practice? (The larger the number, the higher the representativeness of the categories) | 4.0 | 4.0 |
| Q14 | To what extent do you think that distinguishing the dynamic variables into categories can further help log analysis? | 4.2 | 4.0 |
| Q15 | To what extent do you think that, for different specific tasks/requirements, different categories of dynamic variables may have different importance? | 4.5 | 5.0 |
| Q16 | Existing log abstraction technique is used to abstract the dynamic variables of logs and assist in automatic log analysis. If there is an alternative log abstraction tool that can further keep certain categories of dynamic variables (as specified by the user) and abstract the rest, to what extent do you think this alternative tool can be helpful in log analysis? | 4.5 | 5.0 |

**Additional Comments from the Participants (Q17).** We provide an open-ended question to ask the participants if they would like to share some experience or leave some comments related to log analysis.

Some participants provide comments indicating the importance of dynamic variables in logs. For example, two participants commented that:

*"In practice, we would like to pay attention to the specific parameter in a log, while sometimes we can just use the log template to pinpoint the issue. So this may be related to the specific issue. It should be interesting to study the relationship between issues and parameters or templates."*

*"Determining whether a variable is important or not really depends on the task. For example, if we want to find something related to a detected failure, the error code or the identifiable information will be very important."*

There are also some participants who provide examples related to how can dynamic variables help log analysis. More details will be discussed in Section VI. Overall, developers consider that dynamic variables are important, and their importance is related to what information they record and the specific tasks.

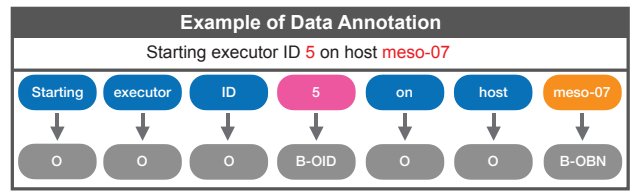

Fig. 3. An example of our log annotation process. Static words are annotated with *O*, object ID is annotated with *B-OID*, and object name is annotated with *B-OBN*.

> We find that different categories of dynamic variables record valuable information that can be important depending on the tasks. We also find that practitioners in our survey consider the distinction of dynamic variables in the process of log abstraction can further help log analysis.

## IV. AN AUTOMATED APPROACH FOR VARIABLE-AWARE LOG ABSTRACTION

Motivated by our empirical findings and practitioners' feedback, in this section, we propose a deep learning based log abstraction approach, called VALB, which is a **V**ariable-**A**ware **L**og a**B**straction technique. Given a set of logs, VALB can identify the static words (i.e., log templates), dynamic variables, and the categories of dynamic variables. Hence, VALB can be used as a general log parser when developers decide to abstract all the categories of dynamic variables. Moreover, as we found in the result of our survey, practitioners consider that different categories of dynamic variables can have different importance depending on the tasks. Therefore, VALB also allows developers to decide which categories of variables they want to keep and preserve the values of such dynamic variables.

We formulate our variable-aware log abstraction process as a sequence tagging problem, which is widely studied in the natural language processing area [39], [40]. A typical usage of sequence tagging in NLP is named entity recognition (e.g., given a sentence, to recognize which word is a person, which word is an organization, etc.). In our study, for a given log message, VALB aims to identify which words are static words, which words are dynamic variables and what are their corresponding categories. Below, we discuss how we annotate the dynamic variables and static words in logs, and the deep learning framework and implementation of VALB.

### A. Data Annotation

VALB is based on supervised deep learning. In order to train the model that can identify the dynamic variables, their corresponding category, and static words in logs, we need to prepare the annotated training data. The training data consists of a specific amount of logs, and each word in the log is annotated with its category. For each word in the log, we use the IOB (inside-outside-beginning) annotation format [41] to annotate it with the categories that we found in Section III. This format uses *"B-"* as the prefix of the beginning word of a named entity and uses *"I-"* as the prefix for the following word (if the following word exists). For the outside word of a named entity, it uses *"O"* as the annotation. In our study, for each dynamic variable, we use the aforementioned prefix as well as the abbreviation illustrated in Table II to annotate the category of each dynamic variable. For each static word, we annotate it as *"O"*.

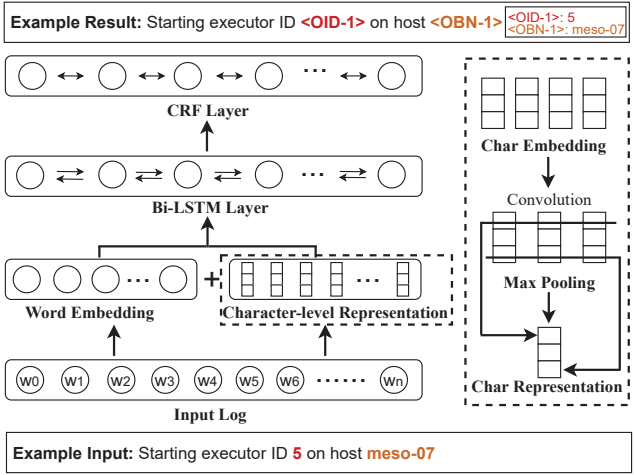

Fig. 4. Overall diagram of our framework. Areas surrounded by dashed lines on the right illustrate the detailed structure for the character-level representation.

Figure 3 shows the annotation process of a simplified log from the Spark data set. In the example, the log message is *"Starting executor ID 5 on host meso-07"*. The words *"Starting"*, *"executor"*, *"ID"*, *"on"*, and *"host"* are static words, so we annotate them as *"O"*. For the word *"5"*, it belongs to the category of *Object ID*, so we annotate it as *"B-OID"*. Similarly, we annotate the word *"meso-07"* as *"B-OBN"* (i.e., *Object Name*).

### B. Deep Learning Framework and Implementation

**Overall Architecture.** Figure 4 shows the overall architecture of our deep learning framework. We first feed the log vectors into an embedding layer. Due to the variety of words in logs (e.g., numbers, normal words, and compound words), we use a combination of word embedding and character-level representation as our embedding layer. The embedding layer learns the relationship among the words and characters in logs and transfers the log vectors into probabilistic representations. We then use a Bi-LSTM (Bidirectional Long Short-Term Memory) layer to model the dependencies among the words in logs. Finally, we use a CRF (Conditional Random Fields) layer to model the relationships among the annotations of categories (e.g., which annotations are likely to appear together) and output the annotation of each word. We then analyze the annotation results and input logs to output the variable-aware log templates as the final results.

**Embedding Layer.** In the embedding layer, we use the concatenation of word embedding and character-level representation to transfer the log vector into probabilistic representations.

*Word Embedding.* For each log in the data set, the word embedding layer captures the relationship among the words in the log and transfers the log into probabilistic representations (i.e., probabilistic vector). Similar words tend to have a close distance in the vector space [42], [43], [44]. In our study, we use the GloVe embeddings [45] which are trained from six billion words collected from Wikipedia and the web.

*Character-level Representation.* Unlike static words that are always constant, the dynamic variables in logs can have various values based on different system behaviors. Moreover, many of the dynamic variables are numeric words (e.g., the category Object Amount in Section III). The simple numbers from 0 to 9 can almost have unlimited potential of creating "new words" depending on various run-time information. This may result in very large size of the vocabulary and the OOV (out-of-vocabulary) problem while applying the models [46]. Hence, we also include character-level representation together with the word embedding layer. The combined embedding layer can catch the relationships among both the words as well as the characters [47], [48]. We train the word embedding from the words in logs and then use CNN (Convolutional Neural Network) with max pooling to capture the relationship among the characters in words and build the character-level representation [49], [50]. We then concatenate the word embedding vector and character-level representation together and feed the combined embedding vector into the next layer.

**Bi-LSTM Layer.** Recurrent Neural Network is powerful at capturing the dependencies in sequential data [51], [52]. Log is a series of words that have sequential dependencies among the words. Similar to sentences in natural language, the words in a log may also have dependencies on past (i.e., words on the left) or future words (i.e., words on the right). Hence, we use Bi-LSTM (Bidirectional Long Short-Term Memory) [40], [53], a variant of RNN to capture the long term dependencies in the words from both directions. We then feed the output vectors in this layer that contains the dependency information in logs to the next layer.

**CRF Layer.** In sequence tagging tasks, the CRF (Conditional Random Fields) layer leverages the past and future tag in a sentence to predict the current tag [54]. It can learn the relationships and dependencies among the resulted annotations (e.g., which annotations are likely to appear together, and which annotations are not). In our deep learning framework, the CRF layer uses the vectors from the Bi-LSTM layer and leverages the category annotations of the past and the future words to predict the annotation (i.e., variable category) of the current word. For each line of log, the CRF layer outputs the category of each word in the log as the final result. We then analyze the results from CRF layer and output the static words (i.e., annotated as "O") and categories of dynamic variables as the result of variable-aware log templates (as shown in "Example Result" of Figure 4). The result also includes the value of each dynamic variable, developers can specify the categories and preserve their values.

**Implementation and Hyper-parameters.** We use Tensor-

flow [55] to implement our deep learning framework. To mitigate the impact of overfitting, we apply the dropout method for embedding layer and RNN layer, with a dropout rate of 0.2 [56], [57], [58]. For the embedding layer, we set the dimension as 300, filter size as 50, and kernel size as 3 for character-level embedding and CNN, and set the dimension as 100 for word embedding [59], [40], [46]. For the RNN layer, we set the hidden units as 128 [46]. For the training process, we set the number of epoch as 30 and the batch size as 8 [60], [57], [61].

## V. EVALUATION OF VALB

In this section, we first discuss the experimental setup to evaluate VALB. We then propose three research questions and discuss the results.

### A. Experimental Setup.

**Data Preparation:** We continue to use the log abstraction benchmark data sets provided by the *LogPAI* project [30] (as discussed in Section III), which is widely used by prior log abstraction studies as the evaluation benchmarks [19], [20], [22], to train and evaluate the models. Specifically, we annotate all the 2,000 logs in each of the 16 data sets following the process discussed in Section IV. For each data set, we randomly split the 2,000 logs into training (20%), validation (20%), and testing data sets (60%). The intention of choosing a small size of training and validation data set is that, we want to investigate if training on a small data set can also achieve promising results. If so, the effort of preparing the training data sets can then be mitigated.

**Alternative Approach:** Since there is no prior work on abstracting specific categories of dynamic variables in logs, we use the framework of VALB that excludes the char-level representations (i.e., only using regular word embedding) as the baseline approach to compare with. The purpose is to examine how character-level representation can help to model the diverse lexical usage of dynamic variables.

### B. Research Questions

We discuss the results by answering three research questions. In RQ1, we use VALB as a general log parser that abstracts all the dynamic variables and compare the accuracy with other state-of-the-art log parsers. In RQ2, we examine the accuracy of VALB on variable-aware log abstraction that further identifies the category of dynamic variables. In RQ3, we investigate whether the trained models of VALB can be easily adopted to a new project.

**RQ1: What is the Accuracy of VALB on General Log Abstraction?**

**Motivation.** Prior log abstraction techniques aim at identifying the dynamic parts in the logs and completely abstract them [19]. Similar to prior works, VALB can also be used for general log abstraction if we only identify the dynamic variables and do not consider their categories. In this RQ, we investigate the accuracy of VALB when we use it for general log abstraction and compare it with other state-of-the-arts.

**Approach.** For each data set, we train and validate the model using the training and validation data sets and evaluate the accuracy on the testing data set. When we are training and evaluating the models, we first transfer the annotations of all the categories of dynamic variables to a single annotation that indicates the word is a variable, regardless of their categories. Given a log, VALB can thus identify which words are static words and which words are dynamic variables, and output log templates without dynamic variables as what prior log abstraction works do.

For the accuracy of log abstraction, there are mainly two definitions: 1) a log is considered as correctly parsed if its event template corresponds to the same group of log messages in the ground truth [19], [22]; 2) a log is considered correctly parsed if and only if all of its static words and dynamic variables are correctly identified [20] (the category of dynamic variable is not considered). The first definition of accuracy does not examine if each word is correctly parsed. Therefore, we use the second definition of accuracy to examine the performance of VALB and other works on general log abstraction. The result of accuracy is computed as the ratio of correctly parsed logs against all the parsed logs. We refer to this accuracy as *general accuracy* in the rest of our paper. We use VALB as a general log parser (i.e., abstracting all the identified dynamic variables) and compare the accuracy with the top-3 state-of-the-art log parsers that have the highest accuracy reported in a prior study [20] (i.e., Logram [20], Drain [22], and AEL [23]) as well as our baseline approach.

**Results and Discussions.** Table V presents the general accuracy of our approach (VALB), the baseline (Base), and the other state-of-the-art log parsers. Each number indicates the ratio of the correctly parsed logs. The accuracy that is higher than 90.0% is marked in bold, and the highest accuracy among all the log parsers is marked with a star mark (*). Overall, VALB achieves the best accuracy in 15 out of the 16 data sets and the highest average accuracy across the data sets (96.1%). For the data set that VALB does not achieve the best accuracy, the accuracy of VALB is also close to the highest approach (e.g., VALB achieves an accuracy of 97.0% in HDFS, which is slightly lower than the highest accuracy of 99.9% achieved by AEL and Drain).

> VALB achieves a high accuracy in general log abstraction that abstracts all the identified dynamic variables (96.1% on average), which outperforms other state-of-the-arts.

**RQ2: What is the Accuracy of VALB on Variable-aware Log Abstraction?**

**Motivation.** In our empirical study and survey, we find that practitioners acknowledge the importance of dynamic variables, and different categories of dynamic variables may have different usages depending on the tasks or scenarios. The findings point out the need of a variable-aware log abstraction technique that can preserve the value of specific categories of dynamic variables in the process of log abstraction. In this RQ, we evaluate the accuracy of VALB on variable-aware log

TABLE V
ACCURACY (%) OF VALB ON GENERAL LOG ABSTRACTION COMPARED WITH OTHER LOG PARSERS AND THE BASELINE (Base). **BOLD** NUMBERS: HIGHER THAN 90, STAR MARK (*): HIGHEST ACCURACY IN EACH ROW.

| Dataset | AEL | Drain | Logram | Base | VALB |
|---|---|---|---|---|---|
| **Android** | 86.7 | **93.3** | 84.8 | 79.8 | **93.5*** |
| **Apache** | 69.3 | 69.3 | 69.9 | **91.0** | **100.0*** |
| **BGL** | 81.8 | 82.2 | 74.0 | 83.0 | **91.3*** |
| **Hadoop** | 53.9 | 54.5 | **96.5** | **92.6** | **97.7*** |
| **HDFS** | **99.9*** | **99.9*** | **98.1** | **91.1** | **97.0** |
| **HealthAPP** | 61.5 | 60.9 | **96.9** | 75.8 | **99.3*** |
| **HPC** | **99.0** | **92.9** | **95.9** | **90.8** | **99.2*** |
| **Linux** | 24.1 | 25.0 | 46.0 | **93.8** | **96.5*** |
| **Mac** | 57.9 | 51.5 | 66.6 | 67.2 | **86.6*** |
| **OpenSSH** | 24.7 | 50.7 | 54.5 | **95.8** | **98.2*** |
| **OpenStack** | 71.8 | 53.8 | 84.7 | **92.0** | **93.8*** |
| **Proxifier** | **96.8** | **97.3** | **95.1** | **100.0*** | **100.0*** |
| **Spark** | **96.5** | **90.2** | **90.3** | **91.2** | **99.3*** |
| **Thunderbird** | 78.2 | 80.3 | 76.1 | 83.4 | **88.1*** |
| **Windows** | **98.3** | **98.3** | **95.7** | **91.3** | **99.2*** |
| **Zookeeper** | **92.2** | **96.2** | **95.5** | **92.1** | **98.3*** |
| *Average* | 74.5 | 74.8 | 82.5 | 88.2 | **96.1*** |

TABLE VI
VARIABLE-AWARE ACCURACY (%) OF VALB AND THE BASELINE (Base) DISCUSSED IN RQ2, AND FINE-TUNING MODELS WITH 50 LOGS FROM THE TARGET DATA SET (F-50) DISCUSSED IN RQ3.

| | RQ2 | | RQ3 |
|---|---|---|---|
| **Dataset** | **Base** | **VALB** | **F-50** |
| **Android** | 76.0 | **91.6** | 82.3 |
| **Apache** | **90.5** | **99.3** | **97.0** |
| **BGL** | 82.0 | 89.6 | 86.7 |
| **Hadoop** | **91.8** | **96.8** | **90.1** |
| **HDFS** | 88.9 | **96.5** | **95.0** |
| **HealthAPP** | 75.1 | **98.8** | **92.9** |
| **HPC** | 86.6 | **99.0** | **95.8** |
| **Linux** | **91.6** | **95.9** | **91.0** |
| **Mac** | 63.8 | 86.2 | 78.0 |
| **OpenSSH** | **90.1** | **97.6** | **91.5** |
| **OpenStack** | 89.5 | **93.2** | 88.9 |
| **Proxifier** | **100.0** | **100.0** | **100.0** |
| **Spark** | **90.7** | **99.1** | **92.3** |
| **Thunderbird** | 80.6 | 87.8 | 82.3 |
| **Windows** | **90.4** | **99.0** | **96.7** |
| **Zookeeper** | **91.7** | **98.1** | **95.4** |
| *Average* | 86.2 | **95.5** | **91.0** |

abstraction, as well as the performance on identifying each category of dynamic variables. We study two sub-RQs:

*RQ2-A:* What is the accuracy of VALB on variable-aware log abstraction that can identify the static and dynamic parts in logs, and also further identify the categories of dynamic variables?

*RQ2-B:* What is the performance of VALB on identifying different categories of dynamic variables in logs?

**Approach.** Below, we discuss the approach of each sub-RQ.

*RQ2-A:* Apart from identifying static and dynamic parts in logs (i.e., general log abstraction), VALB can also identify the categories of dynamic variables (i.e., variable-aware log abstraction). To compute the accuracy of variable-aware log abstraction, we consider a log is correctly parsed when: 1) the static and dynamic parts are correctly identified **and** 2) all the categories of dynamic variables in a log are also correctly identified. We refer to this accuracy as *variable-aware accuracy* in the rest of our paper. For each data set, we train and validate the model using the training and validation data sets and evaluate the variable-aware accuracy on the testing data set. Note that since prior log abstraction approaches cannot distinguish the categories of dynamic variables, we only compare the variable-aware accuracy of VALB with the baseline approach.

*RQ2-B:* In this sub-RQ, we further investigate the performance of VALB on identifying each category of dynamic variables. Specifically, we combine the results of all the 16 data sets in RQ2-A and compute an overall precision, recall, and F1 score for each of the 10 categories of the dynamic variables. These metrics are widely used by prior studies on sequence tagging [39], [40]. For each category, precision represents the ability of correctly identifying this category of dynamic variables (i.e., true positive divided by the sum of true positive and false positive); recall represents the ability of how many words in the log that belong to this category can be identified (i.e., true positive divided by the sum of true positive and false negative); and F1 score evaluates if the approach can both accurately and sufficiently identify the words that belong to this category. We also repeat the same process for the baseline and compare the baseline's performance with VALB.

**Results and Discussions.** We present and discuss the results of the two sub-RQs, respectively.

*RQ2-A:* Table VI presents the variable-aware accuracy of VALB and the baseline approach. Overall, VALB achieves a high variable-aware accuracy ranging from 86.2% in Mac to 100.0% in Proxifier, which is also close to the general accuracy as discussed in RQ2-A. The average variable-aware accuracy of VALB is 95.5%, which is higher than the baseline (i.e., 86.2%). The results show that apart from general log abstraction, VALB can also efficiently identify the categories of dynamic variables in the logs to perform a variable-aware log abstraction. Practitioners can specify the categories of dynamic variables based on their needs, and the values of such dynamic variables will be preserved along with the log templates for further log analysis.

*RQ2-B:* Table VII shows the results of identifying different categories of dynamic variables using our approach (VALB) and the baseline (Base). We present the average results on identifying each category of dynamic variables from all the data sets to concisely show the overall performance. Each number represents for the average number computed from all the data sets. The *Average* line shows the arithmetic mean value of the corresponding column. Overall, VALB achieves over 90% in precision, recall, and F1 score for all the categories of dynamic variables and performs better than the baseline. VALB achieves an average precision of 96.2%, an average recall of 96.5%, and an average F1 score of 96.3%; while the baseline achieves 86.6%, 87.8%, and 87.1%, respectively. VALB also has over 99% precision for Object Name (99.8%) and Time or Duration of an Action (99.7%).

TABLE VII
THE RESULTS OF IDENTIFYING DIFFERENT CATEGORIES OF DYNAMIC
VARIABLES BY OUR APPROACH (VALB) AND THE BASELINE (Base).

| Category | Precision (%) VALB | Precision (%) Base | Recall (%) VALB | Recall (%) Base | F1 (%) VALB | F1 (%) Base |
|---|---|---|---|---|---|---|
| **Object ID** | 96.5 | 89.2 | 95.9 | 93.1 | 96.2 | 91.1 |
| **Location Indicator** | 97.1 | 95.2 | 96.3 | 91.3 | 96.7 | 93.2 |
| **Object Name** | 99.8 | 95.5 | 98.3 | 95.8 | 99.0 | 95.6 |
| **Type Indicator** | 92.8 | 74.1 | 95.9 | 67.2 | 94.3 | 70.5 |
| **Switch Indicator** | 96.7 | 87.3 | 98.2 | 83.8 | 97.4 | 85.5 |
| **T. or D. of an Action** | 99.7 | 92.1 | 98.0 | 97.8 | 98.8 | 94.9 |
| **Computing Resources** | 98.7 | 91.2 | 97.3 | 91.7 | 98.0 | 91.4 |
| **Object Amount** | 92.5 | 77.5 | 96.9 | 87.8 | 94.6 | 82.3 |
| **Status Code** | 97.2 | 91.5 | 95.2 | 87.3 | 96.2 | 89.3 |
| **Other Parameters** | 91.1 | 72.9 | 93.0 | 82.2 | 92.0 | 77.2 |
| *Average* | 96.2 | 86.6 | 96.5 | 87.8 | 96.3 | 87.1 |

> VALB can effectively identify the categories of dynamic variables and achieves a high accuracy in variable-aware accuracy (95.5% on average), which outperforms the baseline approach without char-level representations (86.2% on average).

**RQ3: Can the models of VALB be easily leveraged in a new project?**

**Motivation.** Given that VALB is a supervised approach, the effectiveness of the models may rely on the training data. In this RQ, we would like to investigate how generalizable are the models of VALB. Specifically, we study how effective is VALB when the models are trained from other data and fine-tuned with a small size of data in the target data set.

**Approach.** We apply fine-tuning on existing models to investigate if VALB can be easily adopted to a new project with mitigated effort on data preparation. Specifically, for each target data set in the 16 data sets, we combine the training and validation data from the remaining 15 data sets and train a model. We then use a small size of logs (5, 10, 30, 50, and 100) from the training and validation data sets, respectively, from the target data set to fine-tune the model. We further use the fine-tuned model on the target testing data set to examine the variable-aware accuracy and compute an average number by combining the results of all the 16 target data sets together to show an overall trend for different size of fine-tuning logs.

**Results and Discussions.** Figure 5 shows the average variable-aware accuracy of models fine-tuned with different number of logs in the target data set (i.e., F-5, F-10, F-30, F-50, and F-50), models without fine tuning (i.e., F-0), and the original results discussed in RQ2-A (i.e., Original). Overall, the average accuracy increases as the growth of the size of fine-tuning logs in the target data set, from 51.1% without fine-tuning logs to 92.2% with 100 fine-tuning logs. It is worth noting that the average variable-aware accuracy is fairly high when the models are fine-tuned with 50 logs in the target data set (i.e., 91.0% for F-50) and comparable to the original results discussed in RQ2-A.

In the last column of Table VI, we further present the detailed variable-aware accuracy of F-50 for each data set. We find that the fine-tuned models using 50 logs from the target data set (F-50) also achieve a high variable-aware accuracy with an average variable-aware accuracy of 91.0%, which is close to the average accuracy of the original results (i.e., 95.5%). Hence, after using the pre-trained models and a small data set of the target project, the models of VALB can be easily adopted to other projects.

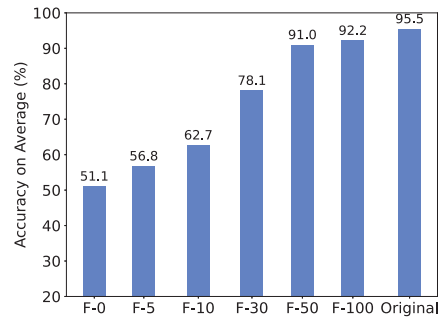

Fig. 5. Average variable-aware accuracy of models fine-tuned with different number of logs in the target data set comparing with the original results in RQ2-A.

> VALB achieves a high variable-aware accuracy using the models fine-tuned with a small amount of data in the target system, and thus can be easily leveraged in a new project.

## VI. DISCUSSION

**Exploring the potential of variable-aware log abstraction on assisting in log-based downstream tasks.** As discussed in Section II, dynamic variables may also contain important information for log analysis tasks, and such information can be preserved using variable-aware log abstraction of VALB. Hence, we further explore how can variable-aware log abstraction help the downstream tasks in log analysis. We conduct our exploration on the log-based anomaly detection benchmark provided by LogPAI [24], [30], which is widely used by other log-based anomaly detection studies [15], [16]. We use the HDFS data set provided by the benchmark to examine the performance of general log abstraction and variable-aware log abstraction on anomaly detection. HDFS data set contains over 11M log messages generated by running Hadoop-based MapReduce jobs on more than 2,000 Amazon's EC2 nodes for 38.7 hours. After grouping the logs with their block ID, there are 575,062 log sequences in total. Around 2.9% of the log sequences indicate anomalies, which are manually labeled by domain experts. We find that the logs in HDFS data set has four categories of dynamic variables: Object ID (OID), Location Indicator (LOI), Computing Resources (CRS), and Object Amount (OBA).

We use the anomaly detection techniques provided by the benchmark [24] with top-5 F1-scores (i.e., Decision Tree, SVM, LR, IM, and Clustering). For each technique, we use log sequences without dynamic variables as what prior log abstraction studies do (i.e., Original), with the value of each category of dynamic variable (i.e., OID, LOI, CRS, and OBA), and with all the values of dynamic variables (i.e., All) as the input data, respectively, to examine their performance on anomaly detection. Note that we further exclude the results of Decision Tree since the Precision, Recall, and F1-score are already nearly perfect (99.8%) for Original, and the results are very similar when using log sequences with each category of dynamic variables (from 99.7% to 99.9%).

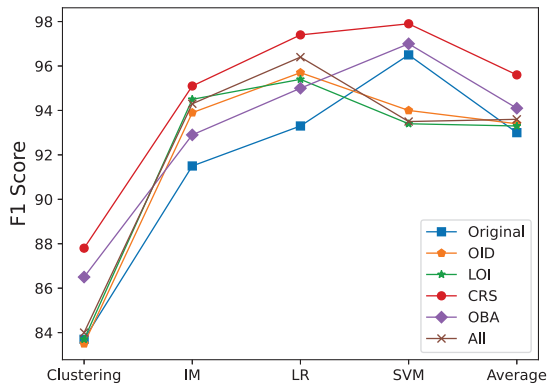

Fig. 6. F1 score achieved by different anomaly detection techniques using sequences of log templates without variables (Original), and using sequences of log templates with corresponding category of variables.

Figure 6 presents the F1 scores achieved by each anomaly detection technique (excluding Decision Tree) using sequences of log templates without variables (i.e., Original), and using sequences of log templates with corresponding category of variables. As we find that the overall trends of Precision and Recall are similar to F1-score, we only present the results of F1-score to have a more concise view. Overall, we find that CRS (i.e., when using the log sequences with dynamic variables of which the category is Computing Resources) achieves the highest F1-score for each of the anomaly detection technique. For other category of variables, there is a fluctuation on the results compared with Original. In short, we find that log sequences with specific categories of dynamic variables (e.g., CRS in this experiment) can improve the performance of log-based anomaly detection.

Apart from anomaly detection, some participants in our survey (as discussed in Section III) also mention some scenarios that dynamic variables in logs can assist in different tasks. For example, one participant comments that:

*"Dynamic variables in the log are very important for parameter tuning works. Especially when the number of parameters is large, using dynamic variables in logs can help to track the performance of each parameter and easy to repeat the best performance."*

The participant mentions that dynamic variables that record the hyper-parameters (e.g., the number of epochs can be represented by the category of *Object Amount* in Section III) can assist in parameter tuning works. Moreover, one participant also mentions that:

*"Some types of variable can be very important for trouble shooting, like the status code. However, it's time-consuming to design regular expressions to grep such variables in each case. It will be helpful to identify such variables without ad-hoc efforts every time."*

Overall, practitioners acknowledge the importance of dynamic variables in practice, and such importance usually depends on the specific tasks. Our study explores the potential of variable-aware log abstraction on assisting in log analysis and sheds light on better leveraging the information in dynamic variables to improve log analysis for future studies.

## VII. THREATS TO VALIDITY

**Construct Validity.** Our approach is based on supervised deep learning, the process of annotation on training and validation data may require extra effort in practice. However, as we discussed in Section V, our approach can achieve promising results when training on small data sets and test on large data sets. Moreover, as we discussed in Section VI, our approach can also achieve encouraging results on the model trained from other projects and fine-tuned on a very small data set (e.g., 50 logs) of the target project. Hence, developers may not need significant time on manually labeling the data.

**Internal Validity.** Regular expression may also be used to extract the dynamic variables in logs, it can work when the format of log messages is clear and easy to distinguish each component in the logs (e.g., structured logs). However, unstructured logs are still very common and may be diverse in terms of their format. It can be difficult to design regular expressions every time. For example, one participant in our survey mentions *"it's time-consuming to design regular expressions to grep variables in each case. It will be helpful to identify such variables without ad-hoc efforts every time."* Our deep learning approach can identify the dynamic variables with more flexibility and mitigate the effort of designing ad-hoc regular expressions. Practitioners can leverage both our approach and regular expressions based on their situations and needs. We conduct manual studies to investigate the characteristic of dynamic variables in logs. To mitigate subjectivity, two of the authors categorize the dynamic variables independently and have a Cohen's Kappa of 0.79, which indicates a substantial level of agreement [38]. Involving third-party experts in log analysis to participate in the manual process may further mitigate this threat.

**External Validity.** We conduct our study on open source log data sets provided by *LogPAI* [30] project. Conducting the study on different log data sets may have different results. For example, new categories of dynamic variables may be derived from other data sets. However, the data sets in *LogPAI* are across various domains and are widely studied by prior log-related studies [20], [16], [22], [15]. Moreover, the categories of dynamic variables are flexible to be updated by leveraging developers' data annotations.

## VIII. RELATED WORKS

**Research on Log Abstraction.** There are many prior studies that propose log abstraction techniques to assist in log analysis. Some prior studies use frequent pattern mining (e.g., SLCT [62], LFA [25], LogCluster [63], Logram [20]) to identify the static words that occur frequently in logs. Some studies leverage clustering algorithms to cluster similar logs (e.g., LKE [64], LogSig [65], SHISO [26], LenMa [27], and LogMine [66]), since logs in the same cluster then tend to have the same log template. Some prior studies use heuristics or combined approaches to identify the static and dynamic parts of logs [22], [23], [28], [29]. For example, Drain [22] uses a fixed-depth tree to maintain log groups with the same log template. IPLoM [28] leverages an iterative partitioning strategy to partition logs into different groups. ULP [29] combines string matching and local frequency analysis to parse large log files. In addition to prior log abstraction techniques that aim to identify and abstract all the dynamic parts in logs, our approach can also distinguish different categories of dynamic variables. Developers can specify the categories variables to keep their values based on needs.

**Deep Learning in Log-related Studies.** Recent studies apply deep learning techniques to address log-related problems. Specifically, those studies are related to logging (i.e., writing logging statements) and log analysis (e.g., anomaly detection). For logging, prior studies proposed deep learning approaches on suggesting variables [46], messages, logging locations, verbosity levels, and a complete logging statement. For logging, deep learning based approaches are used by prior studies to suggest messages [67], variables [46], verbosity levels [68], [61], and the logging locations[69]. For log analysis, Zhang et al. [16] proposed an attention-based Bi-LSTM framework to detect log sequences that have anomalies. Yang et al. [15] used probabilistic label estimation and proposed a semi-supervised anomaly detection framework. Different from prior studies working on logging or log analysis, our study uses deep learning techniques in the process of log abstraction.

IX. CONCLUSION

Log abstraction is an important first step for automated log analysis. Prior log abstraction studies aim to completely abstract the dynamic variables in logs, without considering the great values that dynamic variables may have. Through an empirical study and a survey with industrial practitioners, we find that different categories of the dynamic variables in logs can be important for different tasks, and the distinction of dynamic variables in the process of log abstraction may further help log analysis. We then propose a deep learning based approach, VALB, which can further identify the category of dynamic variables in the process of log abstraction. VALB outperforms state-of-the-art log abstraction techniques on general log abstraction, and also achieves promising results on variable-aware log abstraction. Future studies may investigate the relationship between different categories of dynamic variables and their role in different tasks, in order to better leverage the information recorded in the dynamic variables and further help log analysis.

ACKNOWLEDGEMENTS

Chuan Luo's work is supported by the National Natural Science Foundation of China under Grant 62202025.

REFERENCES

[1] D. Yuan, J. Zheng, S. Park, Y. Zhou, and S. Savage, "Improving software diagnosability via log enhancement," in *ASPLOS '11: Proceedings of the 16th international conference on Architectural support for programming languages and operating systems*, 2011, pp. 3–14.
[2] X. Zhou, X. Peng, T. Xie, J. Sun, C. Ji, D. Liu, Q. Xiang, and C. He, "Latent error prediction and fault localization for microservice applications by learning from system trace logs," in *Proceedings of the ACM Joint Meeting on European Software Engineering Conference and Symposium on the Foundations of Software Engineering, ESEC/SIGSOFT FSE 2019*, 2019, pp. 683–694.
[3] D. Schipper, M. F. Aniche, and A. van Deursen, "Tracing back log data to its log statement: from research to practice," in *Proceedings of the 16th International Conference on Mining Software Repositories, MSR 2019*, 2019, pp. 545–549.
[4] T. Su, L. Fan, S. Chen, Y. Liu, L. Xu, G. Pu, and Z. Su, "Why my app crashes understanding and benchmarking framework-specific exceptions of android apps," *IEEE Transactions on Software Engineering*, 2020.
[5] X. Zhang, Y. Xu, S. Qin, S. He, B. Qiao, Z. Li, H. Zhang, X. Li, Y. Dang, Q. Lin, M. Chintalapati, S. Rajmohan, and D. Zhang, "Onion: identifying incident-indicating logs for cloud systems," in *ESEC/FSE '21: 29th ACM Joint European Software Engineering Conference and Symposium on the Foundations of Software Engineering, Athens, Greece, August 23-28, 2021*, pp. 1253–1263.
[6] A. R. Chen, T.-H. Chen, and S. Wang, "Pathidea: Improving information retrieval-based bug localization by re-constructing execution paths using logs," *IEEE Transactions on Software Engineering*, pp. 2905–2919, 2021.
[7] S. Messaoudi, D. Shin, A. Panichella, D. Bianculli, and L. Briand, "Log-based slicing for system-level test cases," in *2021 ACM SIGSOFT International Symposium on Software Testing and Analysis (ISSTA)*, 2021.
[8] K. Nagaraj, C. E. Killian, and J. Neville, "Structured comparative analysis of systems logs to diagnose performance problems," in *Proceedings of the 9th USENIX Symposium on Networked Systems Design and Implementation*, ser. NSDI '12, 2012, pp. 353–366.
[9] M. Nagappan, K. Wu, and M. A. Vouk, "Efficiently extracting operational profiles from execution logs using suffix arrays," in *ISSRE'09: Proceedings of the 20th IEEE International Conference on Software Reliability Engineering*. IEEE Press, 2009, pp. 41–50.
[10] S. K. Kuttal, A. Sarma, and G. Rothermel, "History repeats itself more easily when you log it: Versioning for mashups," in *2011 IEEE symposium on visual languages and human-centric computing (VL/HCC)*, 2011, pp. 69–72.
[11] J. Cito, P. Leitner, T. Fritz, and H. C. Gall, "The making of cloud applications: An empirical study on software development for the cloud," in *Proceedings of the 2015 10th Joint Meeting on Foundations of Software Engineering*, 2015, pp. 393–403.
[12] D. Gadler, M. Mairegger, A. Janes, and B. Russo, "Mining logs to model the use of a system," in *2017 ACM/IEEE International Symposium on Empirical Software Engineering and Measurement (ESEM)*, 2017, pp. 334–343.
[13] Z. Li, "Towards providing automated supports to developers on writing logging statements," in *Proceedings of the 42nd International Conference on Software Engineering: Companion Proceedings, ICSE 2020*, 2020.
[14] D. Shin, Z. A. Khan, D. Bianculli, and L. Briand, "A theoretical framework for understanding the relationship between log parsing and anomaly detection," in *The 21st International Conference on Runtime Verification*.
[15] L. Yang, J. Chen, Z. Wang, W. Wang, J. Jiang, X. Dong, and W. Zhang, "Plelog: Semi-supervised log-based anomaly detection via probabilistic label estimation," in *43rd IEEE/ACM International Conference on Software Engineering: Companion Proceedings, ICSE Companion 2021, Madrid, Spain, May 25-28, 2021*, 2021, pp. 230–231.
[16] X. Zhang, Y. Xu, Q. Lin, B. Qiao, H. Zhang, Y. Dang, C. Xie, X. Yang, Q. Cheng, Z. Li, J. Chen, X. He, R. Yao, J.-G. Lou, M. Chintalapati, F. Shen, and D. Zhang, "Robust log-based anomaly detection on unstable log data," in *Proceedings of the 2019 27th ACM Joint Meeting on European Software Engineering Conference and Symposium on the Foundations of Software Engineering*, ser. ESEC/FSE 2019, 2019, p. 807–817.
[17] N. Zhao, H. Wang, Z. Li, X. Peng, G. Wang, Z. Pan, Y. Wu, Z. Feng, X. Wen, W. Zhang, K. Sui, and D. Pei, "An empirical investigation of practical log anomaly detection for online service systems," in *ESEC/FSE '21: 29th ACM Joint European Software Engineering Conference and Symposium on the Foundations of Software Engineering, Athens, Greece, August 23-28, 2021*, 2021, pp. 1404–1415.

[18] N. Yang, P. J. L. Cuijpers, R. R. H. Schiffelers, J. Lukkien, and A. Serebrenik, "An interview study of how developers use execution logs in embedded software engineering," in *43rd IEEE/ACM International Conference on Software Engineering: Software Engineering in Practice, ICSE (SEIP) 2021, Madrid, Spain, May 25-28, 2021*, 2021, pp. 61–70.

[19] J. Zhu, S. He, J. Liu, P. He, Q. Xie, Z. Zheng, and M. R. Lyu, "Tools and benchmarks for automated log parsing," in *Proceedings of the 41st International Conference on Software Engineering: Software Engineering in Practice, ICSE (SEIP) 2019, Montreal, QC, Canada, May 25-31, 2019*. IEEE / ACM, 2019, pp. 121–130.

[20] H. Dai, H. Li, C. S. Chen, W. Shang, and T.-H. Chen, "Logram: Efficient log parsing using n-gram dictionaries," *IEEE Transactions on Software Engineering*, 2020.

[21] S. Ma, J. Zhai, Y. Kwon, K. H. Lee, X. Zhang, G. F. Ciocarlie, A. Gehani, V. Yegneswaran, D. Xu, and S. Jha, "Kernel-supported cost-effective audit logging for causality tracking," in *2018 USENIX Annual Technical Conference, USENIX ATC 2018, Boston, MA, USA, July 11-13, 2018*, 2018, pp. 241–254.

[22] P. He, J. Zhu, Z. Zheng, and M. R. Lyu, "Drain: An online log parsing approach with fixed depth tree," in *2017 IEEE international conference on web services (ICWS)*, 2017, pp. 33–40.

[23] Z. M. Jiang, A. E. Hassan, G. Hamann, and P. Flora, "An automated approach for abstracting execution logs to execution events," *J. Softw. Maintenance Res. Pract.*, vol. 20, no. 4, pp. 249–267, 2008.

[24] S. He, J. Zhu, P. He, and M. R. Lyu, "Experience report: System log analysis for anomaly detection," in *2016 IEEE 27th international symposium on software reliability engineering (ISSRE)*. IEEE, 2016, pp. 207–218.

[25] M. Nagappan and M. A. Vouk, "Abstracting log lines to log event types for mining software system logs," in *2010 7th IEEE Working Conference on Mining Software Repositories (MSR 2010)*, 2010, pp. 114–117.

[26] M. Mizutani, "Incremental mining of system log format," in *2013 IEEE International Conference on Services Computing*, 2013, pp. 595–602.

[27] K. Shima, "Length matters: Clustering system log messages using length of words," *arXiv preprint arXiv:1611.03213*, 2016.

[28] A. A. Makanju, A. N. Zincir-Heywood, and E. E. Milios, "Clustering event logs using iterative partitioning," in *Proceedings of the 15th ACM SIGKDD international conference on Knowledge discovery and data mining*, 2009, pp. 1255–1264.

[29] I. Sedki, A. Hamou-Lhadj, O. Ait-Mohamed, and M. A. Shehab, "An effective approach for parsing large log files."

[30] S. He, J. Zhu, P. He, and M. R. Lyu, "Loghub: A large collection of system log datasets towards automated log analytics," *CoRR*, 2020.

[31] E. Kalliamvakou, G. Gousios, K. Blincoe, L. Singer, D. M. German, and D. Damian, "The promises and perils of mining github," in *Proceedings of the 11th working conference on mining software repositories*, 2014, pp. 92–101.

[32] E. Kalliamvakou, C. Bird, T. Zimmermann, A. Begel, R. DeLine, and D. M. German, "What makes a great manager of software engineers?" *IEEE Transactions on Software Engineering*, pp. 87–106, 2017.

[33] Z. Ding, J. Chen, and W. Shang, "Towards the use of the readily available tests from the release pipeline as performance tests," in *Proceedings of the 42nd International Conference on Software Engineering,*, ser. ICSE 2020, 2020.

[34] A. R. Chen, T.-H. Chen, and S. Wang, "Demystifying the challenges and benefits of analyzing user-reported logs in bug reports," *Empirical Software Engineering*, pp. 1–30, 2021.

[35] Z. Li, T. P. Chen, J. Yang, and W. Shang, "DLFinder: characterizing and detecting duplicate logging code smells," in *Proceedings of the 41st International Conference on Software Engineering, ICSE 2019*, 2019, pp. 152–163.

[36] Z. Li, T.-H. Chen, J. Yang, and W. Shang, "Studying duplicate logging statements and their relationships with code clones," *IEEE Transactions on Software Engineering*, pp. 2476–2494, 2021.

[37] G. Bowker and S. L. Star, "Sorting things out," *Classification and its consequences*, vol. 4, 1999.

[38] J. Sim and C. C. Wright, "The kappa statistic in reliability studies: Use, interpretation, and sample size requirements," *Physical Therapy*, vol. 85, no. 3, pp. 257–268, March 2005.

[39] R. Collobert, J. Weston, L. Bottou, M. Karlen, K. Kavukcuoglu, and P. Kuksa, "Natural language processing (almost) from scratch," *Journal of machine learning research*, pp. 2493–2537, 2011.

[40] Z. Huang, W. Xu, and K. Yu, "Bidirectional LSTM-CRF models for sequence tagging," *CoRR*, vol. abs/1508.01991, 2015.

[41] L. A. Ramshaw and M. P. Marcus, "Text chunking using transformation-based learning," in *Natural language processing using very large corpora*, 1999, pp. 157–176.

[42] M. Tufano, C. Watson, G. Bavota, M. Di Penta, M. White, and D. Poshyvanyk, "Deep learning similarities from different representations of source code," in *2018 IEEE/ACM 15th International Conference on Mining Software Repositories (MSR)*, 2018, pp. 542–553.

[43] F. Liu, G. Li, Y. Zhao, and Z. Jin, "Multi-task learning based pre-trained language model for code completion," in *Proceedings of the 35th IEEE/ACM International Conference on Automated Software Engineering*, 2020, pp. 473–485.

[44] H. Wang, X. Xia, D. Lo, Q. He, X. Wang, and J. Grundy, "Context-aware retrieval-based deep commit message generation," *ACM Trans. Softw. Eng. Methodol.*, vol. 30, no. 4, pp. 56:1–56:30, 2021.

[45] J. Pennington, R. Socher, and C. D. Manning, "Glove: Global vectors for word representation," in *Proceedings of the 2014 Conference on Empirical Methods in Natural Language Processing, EMNLP 2014, October 25-29, 2014, Doha, Qatar, A meeting of SIGDAT, a Special Interest Group of the ACL*. ACL, 2014, pp. 1532–1543.

[46] Z. Liu, X. Xia, D. Lo, Z. Xing, A. E. Hassan, and S. Li, "Which variables should i log?" *IEEE Transactions on Software Engineering*, 2019, early Access.

[47] C. Dos Santos and B. Zadrozny, "Learning character-level representations for part-of-speech tagging," in *International Conference on Machine Learning*. PMLR, 2014, pp. 1818–1826.

[48] J. P. Chiu and E. Nichols, "Named entity recognition with bidirectional lstm-cnns," *Transactions of the Association for Computational Linguistics*, pp. 357–370, 2016.

[49] J. Chen, S. Zhang, X. He, Q. Lin, H. Zhang, D. Hao, Y. Kang, F. Gao, Z. Xu, Y. Dang *et al.*, "How incidental are the incidents? characterizing and prioritizing incidents for large-scale online service systems," in *Proceedings of the 35th IEEE/ACM International Conference on Automated Software Engineering*, 2020, pp. 373–384.

[50] A. Nikanjam, H. B. Braiek, M. M. Morovati, and F. Khomh, "Automatic fault detection for deep learning programs using graph transformations," *TOSEM*, 2021.

[51] Y. Xu, L. Mou, G. Li, Y. Chen, H. Peng, and Z. Jin, "Classifying relations via long short term memory networks along shortest dependency paths," in *Proceedings of the 2015 conference on empirical methods in natural language processing*, 2015, pp. 1785–1794.

[52] A. Mazuera-Rozo, A. Mojica-Hanke, M. Linares-Vásquez, and G. Bavota, "Shallow or deep? an empirical study on detecting vulnerabilities using deep learning," in *29th IEEE/ACM International Conference on Program Comprehension, ICPC 2021, Madrid, Spain, May 20-21, 2021*. IEEE, 2021, pp. 276–287.

[53] B. D. Q. Nghi, Y. Yu, and L. Jiang, "Bilateral dependency neural networks for cross-language algorithm classification," in *26th IEEE International Conference on Software Analysis, Evolution and Reengineering, SANER 2019*, 2019, pp. 422–433.

[54] J. D. Lafferty, A. McCallum, and F. C. N. Pereira, "Conditional random fields: Probabilistic models for segmenting and labeling sequence data," in *Proceedings of the Eighteenth International Conference on Machine Learning (ICML 2001), Williams College, Williamstown, MA, USA, June 28 - July 1, 2001*, 2001, pp. 282–289.

[55] "Tensorflow: An end-to-end open source machine learning platform," https://www.tensorflow.org/, last checked Aug. 2021.

[56] N. Srivastava, G. E. Hinton, A. Krizhevsky, I. Sutskever, and R. Salakhutdinov, "Dropout: a simple way to prevent neural networks from overfitting," *J. Mach. Learn. Res.*, vol. 15, no. 1, pp. 1929–1958, 2014.

[57] T. Hoang, H. K. Dam, Y. Kamei, D. Lo, and N. Ubayashi, "Deepjit: an end-to-end deep learning framework for just-in-time defect prediction," in *Proceedings of the 16th International Conference on Mining Software Repositories, MSR 2019*, 2019, pp. 34–45.

[58] M. Wardat, W. Le, and H. Rajan, "Deeplocalize: Fault localization for deep neural networks," in *43rd IEEE/ACM International Conference on Software Engineering, ICSE 2021, Madrid, Spain, 22-30 May 2021*. IEEE, 2021, pp. 251–262.

[59] J. Zhang, X. Wang, H. Zhang, H. Sun, K. Wang, and X. Liu, "A novel neural source code representation based on abstract syntax tree," in *Proceedings of the 41st International Conference on Software Engineering, ICSE 2019*, 2019, pp. 783–794.

[60] T. Zhang, C. Gao, L. Ma, M. R. Lyu, and M. Kim, "An empirical study of common challenges in developing deep learning applications," in *30th IEEE International Symposium on Software Reliability Engineering, ISSRE 2019*, 2019, pp. 104–115.

[61] Z. Li, H. Li, T.-H. P. Chen, and W. Shang, "Deeplv: Suggesting log levels using ordinal based neural networks," in *2021 IEEE/ACM 43rd International Conference on Software Engineering (ICSE)*. IEEE, 2021, pp. 1461–1472.

[62] R. Vaarandi, "A data clustering algorithm for mining patterns from event logs," in *Proceedings of the 3rd IEEE Workshop on IP Operations & Management (IPOM 2003)(IEEE Cat. No. 03EX764)*, 2003, pp. 119–126.

[63] R. Vaarandi and M. Pihelgas, "Logcluster-a data clustering and pattern mining algorithm for event logs," in *2015 11th International conference on network and service management (CNSM)*, 2015, pp. 1–7.

[64] Q. Fu, J.-G. Lou, Y. Wang, and J. Li, "Execution anomaly detection in distributed systems through unstructured log analysis," in *2009 ninth IEEE international conference on data mining*, 2009, pp. 149–158.

[65] L. Tang, T. Li, and C.-S. Perng, "Logsig: Generating system events from raw textual logs," in *Proceedings of the 20th ACM international conference on Information and knowledge management*, pp. 785–794.

[66] H. Hamooni, B. Debnath, J. Xu, H. Zhang, G. Jiang, and A. Mueen, "Logmine: Fast pattern recognition for log analytics," in *Proceedings of the 25th ACM International on Conference on Information and Knowledge Management*, pp. 1573–1582.

[67] Z. Ding, H. Li, and W. Shang, "Logentext: Automatically generating logging texts using neural machine translation," in *2022 IEEE International Conference on Software Analysis, Evolution and Reengineering (SANER)*, pp. 349–360.

[68] J. Liu, J. Zeng, X. Wang, K. Ji, and Z. Liang, "Tell: log level suggestions via modeling multi-level code block information," in *Proceedings of the 31st ACM SIGSOFT International Symposium on Software Testing and Analysis*, 2022, pp. 27–38.

[69] Z. Li, T. Chen, and W. Shang, "Where shall we log? studying and suggesting logging locations in code blocks," in *35th IEEE/ACM International Conference on Automated Software Engineering, ASE 2020*, 2020, pp. 361–372.